# Evolution of anisotropic magnetostriction in $LaMn_{1-x}Co_xO_3$ ($0.1 \leq x \leq 0.9$)

M. Manikandan and R. Mahendiran*

*Department of Physics, National University of Singapore, 2 Science Drive 3,*
*Singapore 117551, Republic of Singapore*

## Abstract

Polycrystalline samples of $LaMn_{1-x}Co_xO_3$ series over a wide compositional range ($0.1 \leq x \leq 0.9$) were synthesized by microwave irradiation of oxide precursors and their magnetic and magnetostrictive properties were investigated. Magnetostrictions parallel ($\lambda_{par}$) and perpendicular ($\lambda_{per}$) to the applied magnetic field were measured to estimate anisotropic ($\lambda_t$) and volume ($\lambda_\omega$) magnetostrictions. In all the compositions, $\lambda_{par}$ ($\lambda_{per}$) is negative (positive) and $\lambda_{t,} >> \lambda_{\omega,}$ suggesting dominance of anisotropic lattice distortion under a magnetic field. The value of $\lambda_t$ at 10 K for $H$ = 50 kOe initially increases with $x$ from 178 ppm for $x$ = 0.1 to a maximum of 1221 ppm at $x$ = 0.5 before decreasing for higher $x$. The composition dependence of magnetostriction is asymmetric about $x$ = 0.5, the decrease for $x$ > 0.5 is more steeper than for $x$ < 0.5 whereas saturation magnetization decreases monotonically with increasing $x$ except for an abrupt change at $x$ = 0.6. The largest anisotropic magnetostriction observed for $x$ = 0.5 is attributed to the presence of high-spin $Co^{2+}$ ions with a non-zero orbital moment in the maximum fraction whereas $Mn^{3+/4+}$ or $Co^{3+}$ ions play minor roles. The composition dependence of magnetostriction is suggested to arise from changes in the structure, valence states of Mn and Co ions and magnetic interactions among them.

*Author for correspondence: R. Mahendiran (email: phyrm@nus.edu.sg)

## I. INTRODUCTION

In contrast to ferromagnetism and metallicity induced by substituting a divalent alkaline cation (e.g., $Sr^{2+}$) for $La^{3+}$ cation in $LaMnO_3$ and $LaCoO_3$, the low-temperature phase of $LaMn_{1-x}Co_xO_3$ transits from an antiferromagnetic insulator ($x$ = 0) to a non-magnetic ($x$ = 1) insulator through a ferromagnetic insulator at the intermediate composition $LaMn_{0.5}Co_{0.5}O_3$ [1]. These changes in the magnetic and electrical properties arise from modifications in valences of Mn and Co ions, spin-state of Co ion, and magnetic interactions between them. While both transition metal cations (Mn and Co) in the end compounds of the $LaMn_{1-x}Co_xO_3$ series are trivalent, they adopt 4+ and 2+ valence states, respectively, in $LaMn_{0.5}Co_{0.5}O_3$, and order next to each other in a rock salt configuration, leading to a double perovskite structure ($La_2MnCoO_6$) [2,3]. X-ray absorption and magnetic dichroism studies support the presence of high-spin $Co^{2+}(d^7:t_{2g}^5e_g^2)$ with an appreciable orbital moment [4,5]. The degree of B-site ordering is influenced by final sintering temperature, rate of cooling, and oxygen partial pressure applied during or post-synthesis [6,7,8,9]. Slow cooling enhances B-site cation ordering, resulting in higher saturation magnetization and ferromagnetic Curie temperature [10,11,12]. Ferromagnetism in $La_2CoMnO_6$ is understood as a result of positive superexchange interaction between $Co^{2+}$ ($S$ = 3/2) and $Mn^{4+}$ ($S$ = 3/2) ions [2]. Ferromagnetic transition temperature in the $LaMn_{1-x}Co_xO_3$ series reaches a maximum value of 225±2 K for $x$ = 0.5 [1-12].

Observations of magnetic field tuneable dielectric permittivity [13,14,15,16], large negative magnetoresistance (~80% in a magnetic field of $H$ = 80 kOe at 5 K) [17], and tensile strain-induced perpendicular magnetic anisotropy in thin film form [18], have revived interest in $La_2CoMnO_6$. Interestingly, polycrystalline $La_2CoMnO_6$ also exhibits a giant negative magnetostriction ($\lambda_{par}$ = -610 ppm in a field of 50 kOe at 10 K) along the direction of the applied *dc* magnetic field, which was suggested to lattice contraction caused by reorientation of a non-degenerate $t_{2g}$ orbital along with the rotation of magnetization due to strong spin-orbit coupling associated with the $Co^{2+}$: $t_{2g}^5e_g^2$ ion [19]. This magnetostriction value is larger than that exhibited by spinel ferrite $CoFe_2O_4$ possessing divalent cobalt ions [20]. More recently, Boldrin *et al.*, [21] also report a high longitudinal magnetostriction (~400 ppm for 50 kOe) in $La_2CoMnO_6$ and suggest that magnetoelectricity in this compound is driven by magnetostriction and modulation of orbital hybridization by the applied magnetic field. Magnetostriction in cobalt oxides exhibits many interesting behaviours depending on the valence and spin states of Co-ions. For example, the parent compound $LaCoO_3$ shows an abrupt volume expansion above 600 kOe at 4.2 K due to a field-induced low-spin ($S = 0$) to high-spin ($S = 2$) transition of $Co^{3+}$ions [22], whereas anisotropic magnetostriction dominates over volume magnetostriction in $La_{1-x}Sr_xCoO_3$ ($x = 0.3$-$0.5$) series hosting mixed valence cobalt ($Co^{3+/4+}$) ions and ferromagnetism [23]. The anisotropic magnetostriction was suggested to arise from orbital instability of $Co^{3+}$ ions following a field-induced low- to intermediate spin-state ($S = 1$) transition. In the polar ferrimagnet $CaBaCo_4O_7$, magnetic field-induced electrical polarization is accompanied by a giant longitudinal negative magnetostriction (≈ -1500 ppm for $H = 80$ kOe at 62 K) [24] and it is suggested to field-induced orbital instability of $Co^{3+}/Co^{2+}$ ions. X-ray absorption spectroscopy studies in $LaMn_{1-x}Co_xO_3$ series indicate that the charge transfer reaction $Co^{3+}$-$Mn^{3+} \leftrightarrow Co^{2+}$-$Mn^{4+}$ gradually shifts to the right as Co content increases for $0.15 < x \leq 0.5$ [25,26,27,28,29,30]. Hence, it is of interest to investigate how the sign and the magnitude of magnetostriction vary with Co content ($x$) in the $LaMn_{1-x}Co_xO_3$ series. Longitudinal magnetostriction (magneto-strain measured along the field direction) alone was measured in the earlier two studies on $La_2CoMnO_6$ [20,22]. In this work, both the longitudinal and the transverse magnetostrictions were measured to estimate the anisotropic and volume contributions as $x$ increases from 0.1 to 0.9 in $LaMn_{1-x}Co_xO_3$ series.

We adopt microwave irradiation of oxide precursors to synthesize $LaMn_{1-x}Co_xO_3$. Microwave-assisted synthesis of transition metal oxides is an emerging area of research. It has the advantage that powder samples can be synthesized within a few seconds to a few minutes depending on the ability of starting chemicals (raw oxides or their hydroxides/carbonates) to absorb microwave radiation [31,32,33]. Microwave (MW) energy is directly converted into heat within the starting chemicals when reorientation of dielectric/magnetic dipoles in them is unable to keep in phase with the oscillating electric/magnetic field component of the impinging microwave radiation. Hence, chemicals with high dielectric/magnetic loss at the MW frequency (usually at 2.4 GHz) heat up rapdily. Since microwave heating is volumetric, dense pellets can be obtained with 10 to 30 minutes of sintering. Our previous work showed that $La_2CoMnO_6$ synthesized by microwave irradiation of oxide precursors, exhibits ferromagnetic Curie temperature and saturation magnetization values comparable to a sample obtained via solid-state reaction and sintering for 12 hours using a conventional resistive heating furnace [20].

## II. Experimental details

Powders of $La_2O_3$, $Mn_2O_3$, and $Co_3O_4$ mixed in a stoichiometric ratio were subjected to microwave irradiation in a multi-mode MW box furnace (Milestone PYRO) operating at a frequency of 2.45 GHz. $La_2O_3$ was pre-heated to 900 ºC for 5 hours in an electrical furnace to remove carbonate and impurities before mixing. The mixed powder was heated to 1200 ºC in 15 minutes using 1600 W microwave power and soaked at that temperature for 30 minutes before switching off the MW power. The powder began cooling soon after the MW power was switched off and reached room temperature in 100 minutes. The mixed powder was then homogeneously ground and pressed into disc-shaped pellets using a hydraulic press. The pellets were again exposed to microwave irradiation for 15 minutes to reach 1200 ºC followed by sintering for 30 minutes. The obtained samples were characterized by powder X-ray diffraction with Cu-$K_{\alpha 1}$ radiation of wavelength 1.5406 Å. Magnetization and magnetostriction were measured in a Physical Property Measuring System (PPMS) using a vibrating sample magnetometer and capacitance dilatometer probes, respectively. Magnetostriction along the direction of the applied *dc* magnetic field ($\lambda_{par}$) and in a transverse direction ($\lambda_{per}$) were measured. A polished cubic sample of size 2 x 2 x 2 mm is placed between two electrodes, one fixed and the other movable. As the sample expands or contracts under an applied magnetic field, displacement of the movable electrode changes the capacitance of the cell that is measured by a high-resolution capacitance bridge (Andeen-Hagerling, model AH2500A)

## III. Results and Discussion

### A. Structure

The room-temperature powder X-ray diffraction patterns of $LaMn_{1-x}Co_xO_3$ for $x$ = 0.2-0.8 along with their Rietveld refinement profiles, are shown in Fig. 1(a). All samples can be indexed within perovskite-type structures without detectable impurity phases, confirming the formation of single-phase compounds. The refined profiles show excellent agreement between the observed (black symbols) and calculated patterns (red solid lines), with low residuals in the difference curves (blue lines), confirming the reliability of the structural models. The refinement results reveal a clear composition-dependent structural evolution. The structural evolution is presented separately in an enlarged view of the maximum-intensity peaks in Fig. 1(b). The compositions $x$ = 0.2 and 0.8 crystallize in a rhombohedral ($R\overline{3}c$) structure, whereas $x = 0.4$ and 0.5 stabilize in a monoclinic ($P2_1/n$) phase. The $x = 0.6$ sample exhibits a mixed-phase behaviour, consisting of both monoclinic and rhombohedral (pseudo-monoclinic) contributions, suggesting a structural transition region. We have listed the lattice parameters in Table 1 (supplementary). Earlier structural studies on samples synthesis via solid-reaction route and sintered at $T$ = 1200-1350 ºC in air using conventional electrical furnace resulted in orthorhombic ($Pbmn$) for $x \leq 0.5$, a mixed phase of orthorhombic and rhombohedral for $x$ = 0.6-0.7 and rhombohedral ($R\overline{3}c$) for $x$ = 0.8-1.0 [34]. Reports suggest that $x$ = 0.5 and even ≈ 0.33 can be fitted better with monoclinic ($P2_1/n$) structure [7,35]. The crystallographic structure around $x$ = 0.5 is sensitive to annealing temperature, atmosphere, and oxygen content [36]. Microwave irradiation promotes an orthorhombic to a rhombohedral structural transition as the microwave power is increased from 1000 W to 1600 W [37]. The sample irradiated at 1600 W showed a paramagnetic-ferromagnetic transition at 240 K accompanied by an insulator-metal transition unlike the insulating antiferromagnetic $LaMnO_3$ synthesised via the conventional heating method. The coexistence of rhombohedral and monoclinic phases at $x$ = 0.6 indicates a competing structural instability, which is likely driven by a delicate balance among lattice strain, electronic configuration, and magnetic interactions.

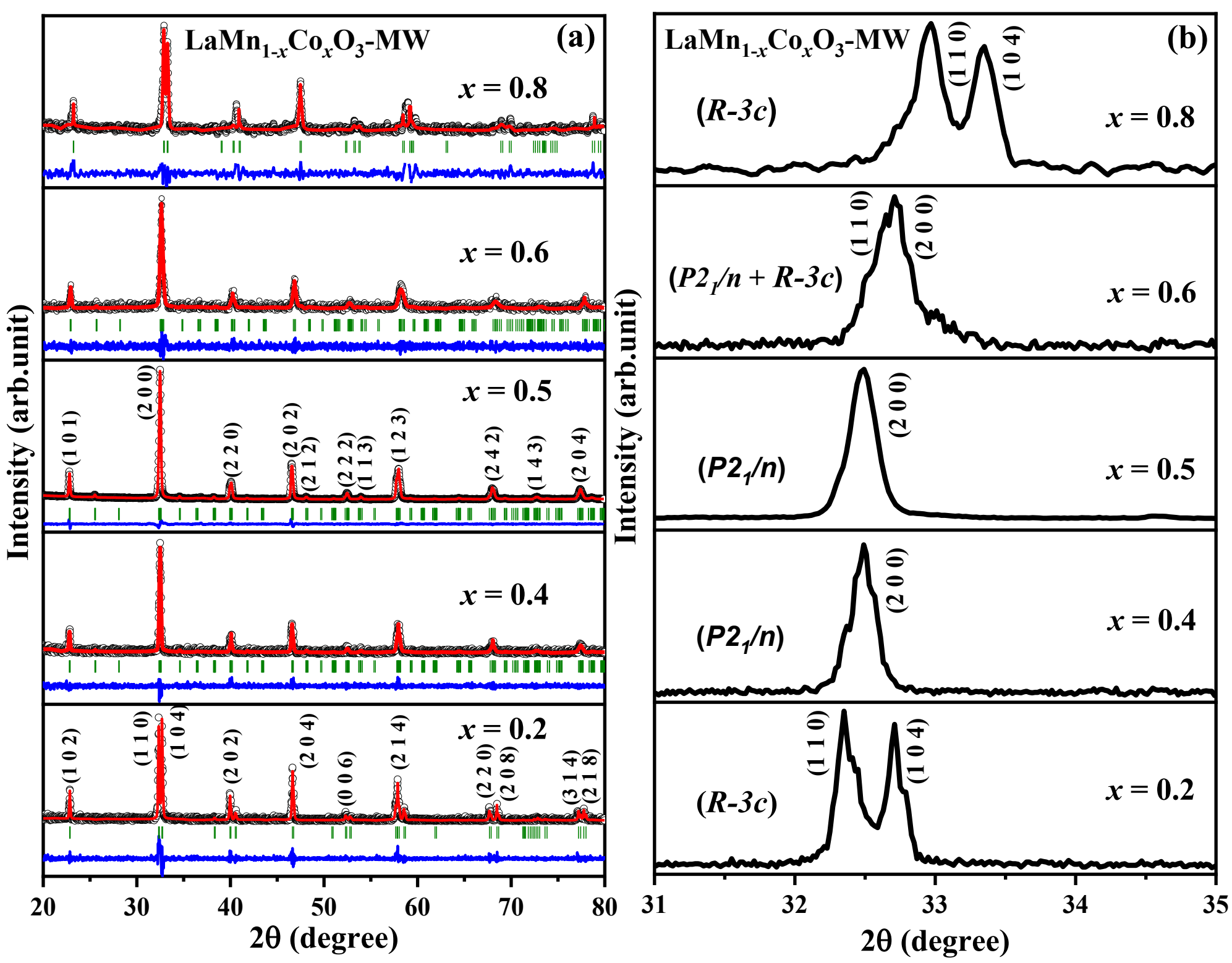

**Fig. 1. (a)** Room-temperature powder X-ray diffraction patterns and corresponding Rietveld refinement profiles of $LaMn_{1-x}Co_xO_3$ ($x$ = 0.2–0.8). The observed data (black symbols),

calculated patterns (solid lines), Bragg reflection positions (Green vertical lines), and difference curves (blue bottom lines) are shown. **(b)** A composition-dependent structural evolution from rhombohedral to monoclinic phase is evident in the enlarged maximum-intensity peaks.

## B. Magnetization

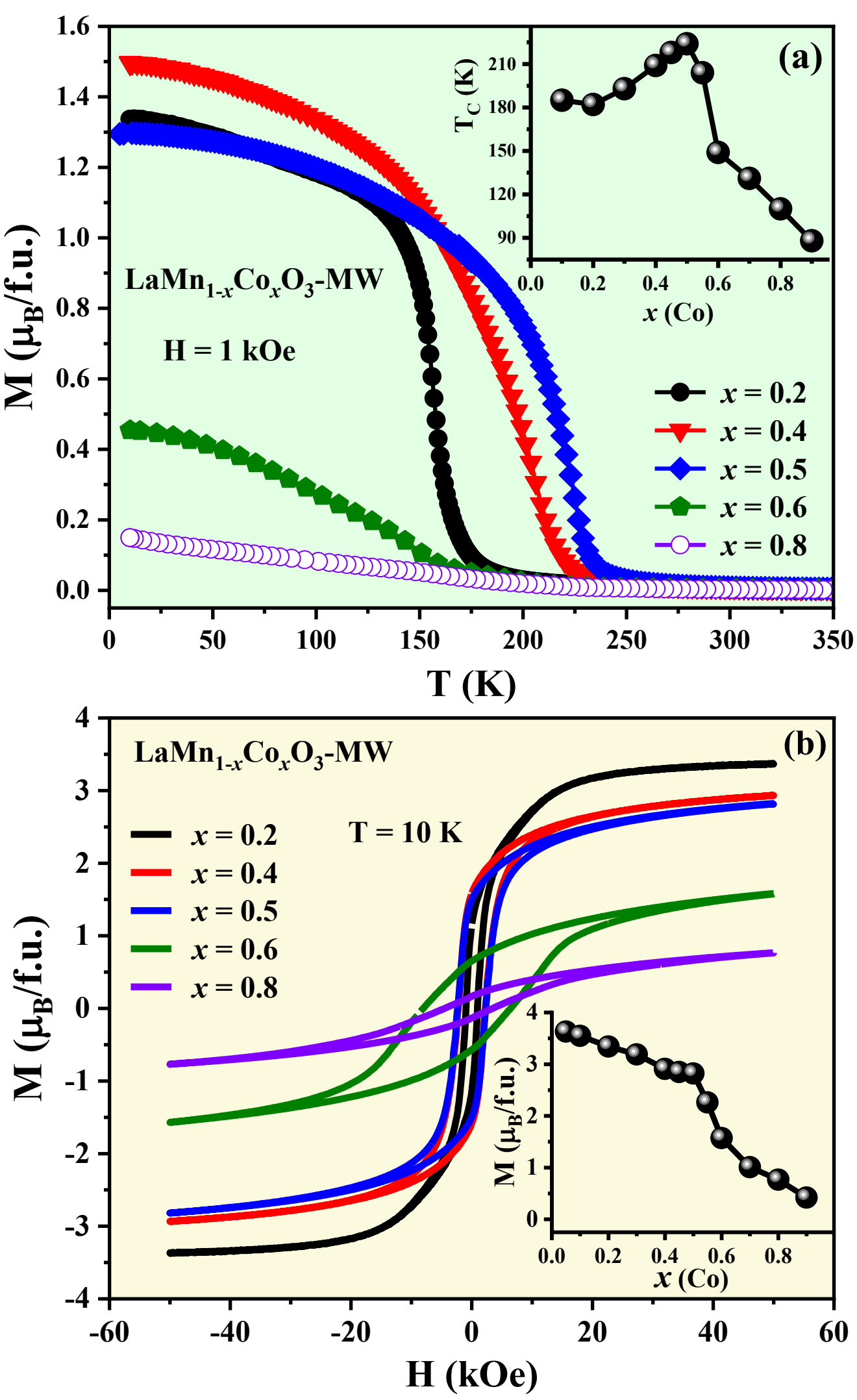

**Fig. 2. (a)** Temperature dependence of magnetization ($M$) under a magnetic field of $H = 1$ kOe and **(b)** $M(H)$ isotherms at 10 K for selected compositions ($x$) in $LaMn_{1-x}Co_xO_3$ ($x = 0.2$ to 0.8). Insets: **(a)** Composition dependence of Curie temperature ($T_C$) and **(b)** Composition dependence of $M$ value at 50 kOe at 10 K.

Figure 2(a) shows the temperature dependence of magnetization ($M$) of $LaMn_{1-x}Co_xO_3$ ($x$ = 0.2–0.8) measured under $H$ = 1 kOe. Samples with $x$ = 0.2, 0.4, and 0.5 exhibit a well-defined ferromagnetic transition characterized by a sharp increase in $M$ at the Curie temperature ($T_C$). However, the magnetic phase transition becomes very broad for $x \geq 0.6$ and the magnitude of magnetization diminishes dramatically. The inset of Fig. 2(a) shows the variation of $T_C$ obtained from the inflexion point of $dM/dT$ with Co concentration. $T_C$ initially decreases from 185 K for $x = 0.1$ to 182 K for $x = 0.2$, then increases to a maximum of 225 K in $x = 0.5$. $T_C$ then decreases slightly for $x = 0.55$ and rapidly above 0.55. This non-monotonic behaviour reflects competition between ferromagnetic double-exchange interactions between $Mn^{3+}$-$Mn^{4+}$ for $x \leq 0.2$, ferromagnetic superexchange interaction between $Co^{2+}$-$Mn^{4+}$ pairs for $0.3 \leq x \leq 0.55$, and antiferromagnetic interactions among like-charged neighbouring ions above $x = 0.55$.

The main panel of Fig. 2(b) illustrates $M(H)$ isotherms at 10 K for $x = 0.2$ to 0.8. Samples with $x \leq 0.5$ show well-defined hysteresis loops with a tendency to saturation at higher fields, confirming robust ferromagnetic behavior. However, the hysteresis loops are more widened at $x = 0.6$ and 0.8. The value of magnetization at 50 kOe decreases with increasing Co content, the decrease being sharper just above $x = 0.55$ as shown in the inset of Fig. 2(b). The saturation value of $M$ in $x = 0.5$ is 2.82 $\mu_B$/f.u. which is close but lower than 3 $\mu_B$/f.u. expected for full ferromagnetic alignment of ordered $Co^{2+}$(S = 3/2) and $Mn^{4+}$(S = 3/2) spins. This deviation could arise from some $Co^{2+}$ ions occupying $Mn^{4+}$ site or vice versa in the double perovskite structure which promotes antiferromagnetic interaction along like charges. The maximum value of M decreases further to 2.26 $\mu_B$/f.u. for $x = 0.55$, to 1.57 $\mu_B$/f.u. for $x = 0.6$, and to 0.89 $\mu_B$/f.u. for $x = 0.8$. This is most likely due to changes in the valences of Mn and Co ions and magnetic interactions between them. On the Mn-rich side, $x = 0.1$ sample shows the maximum saturation magnetization value (3.55 $\mu_B$/f.u.) in the series because of the presence of about 33 % $Mn^{4+}$ ions with interact ferromagnetically with $Mn^{3+}$ ions via the Zener's double exchange interaction [3-7]. Stoichiometric $LaMnO_3$ containing only $Mn^{3+}$ ions (obtained by solid state reaction at T > 1200 °C and annealed in oxygen reducing atmosphere [1]) is orthorhombic at room temperature and an antiferromagnetic insulator ($T_N = 140$ K) whereas creation of 25-40% $Mn^{4+}$ ions by cations deficiency (oxygen excess) results in a rhombohedral structure in which ferromagnetism is mediated by Zener's double exchange interaction between $Mn^{4+}$ and $Mn^{3+}$ ions [38]. Our preliminary report indicates that $LaMnO_{3+\delta}$ sample synthesized by microwave irradiation of 1600 W was rhombohedral with a Curie temperature of 240 K [36]. Magnetic and magnetostriction parameters are summarized in Table 2 of the supplemental.

### C. Magnetostriction

Figures 3(a-d) illustrate the magnetic field dependence of magnetostriction at 10 K in $LaMn_{1-x}Co_xO_3$ ($x$ = 0.2-0.8). Fig. 3(a) and (b) show the longitudinal ($\lambda_{par}$) and transverse ($\lambda_{per}$) magnetostriction measured parallel and perpendicular to the applied magnetic field, respectively, while Fig. 3(c) and (d) show the derived anisotropic ($\lambda_t = \lambda_{par} - \lambda_{per}$) and volume ($\lambda_\omega = \lambda_{par} + 2\lambda_{per}$) magnetostrictions. $\lambda_{par}$ exhibits negative magnetostriction (length contracts along the direction of the applied magnetic field) whereas $\lambda_{per}$ shows positive magnetostriction (length elongates perpendicular to the direction of the applied magnetic field) for all the samples. The magnetostriction curves are symmetric about zero field. While $\lambda_{par}$ and $\lambda_{per}$ for $x$ = 0.2 increase in magnitude at low fields ($H < \pm 10$ kOe) and show a tendency to saturate at higher fields, they increase continuously up to the maximum field for $x \geq 0.3$. The $x$ = 0.6 sample shows relatively broader hysteresis reflecting the trend of hysteresis in magnetization, while it is negligible in $x$ = 0.8. The slope of $\lambda_{par}$($\lambda_{per}$) also changes drastically with $x$ increasing above 0.5. The $\lambda_t$ and $\lambda_\omega$ replicate the features of $\lambda_{par}$ and $\lambda_{per}$ at respective compositions. In all the samples, $\lambda_t > \lambda_\omega$. In $x$ = 0.5, $\lambda_t$ is 1221 ppm at 50 kOe which is nearly three times larger than the volume magnetostriction ($\lambda_\omega$ = 309 ppm) which indicates that anisotropic distortion dominates the lattice.

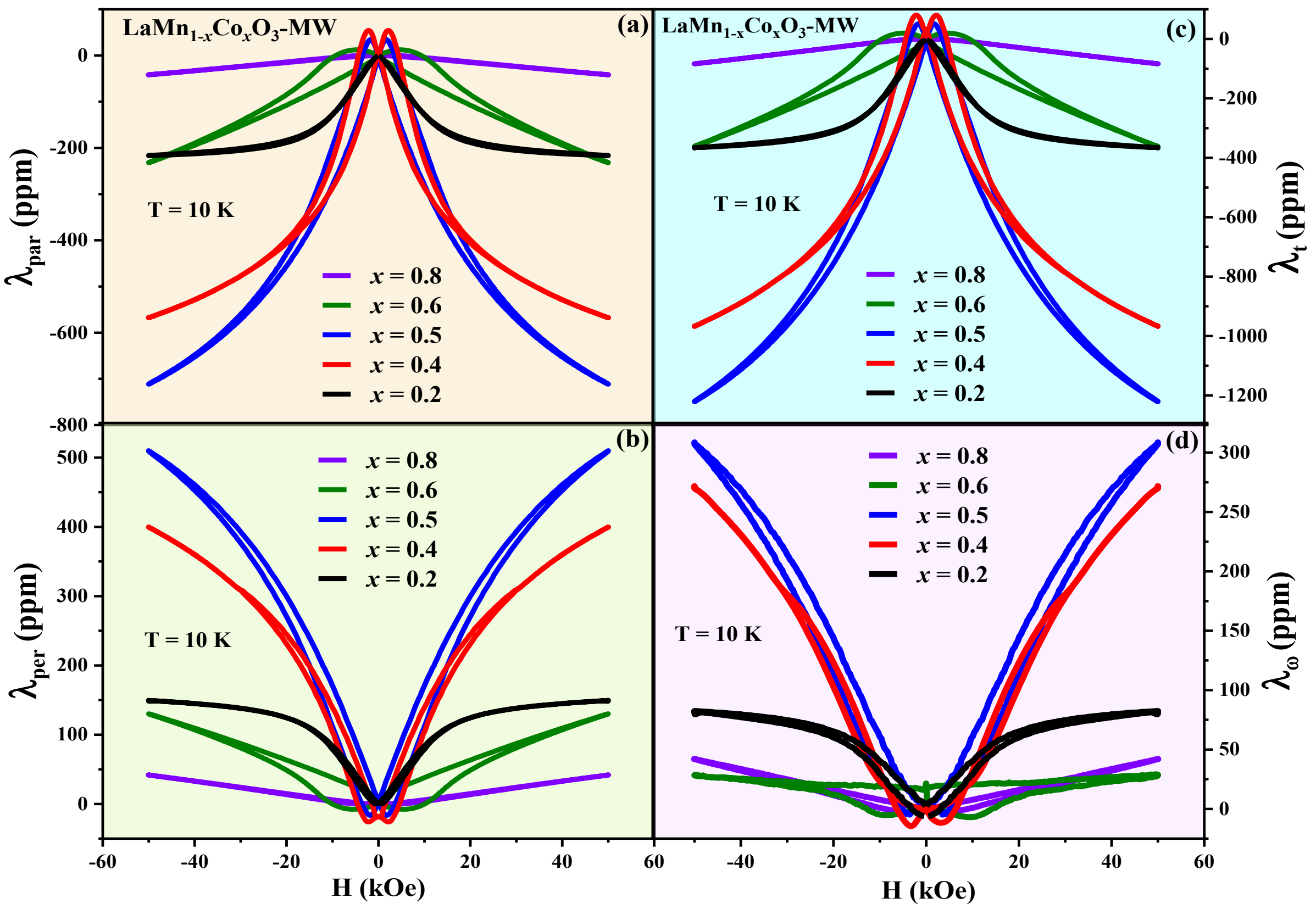

**Fig. 3.** Field dependence of magnetostriction in $LaMn_{1-x}Co_xO_3$ at 10 K for different Co concentrations ($x$ = 0.2, 0.4, 0.5, 0.6, and 0.8), **(a)** parallel ($\lambda_{par}$), **(b)** perpendicular ($\lambda_{per}$), **(c)** anisotropy ($\lambda_t$) and **(d)** volume ($\lambda_\omega$) magnetostrictions.

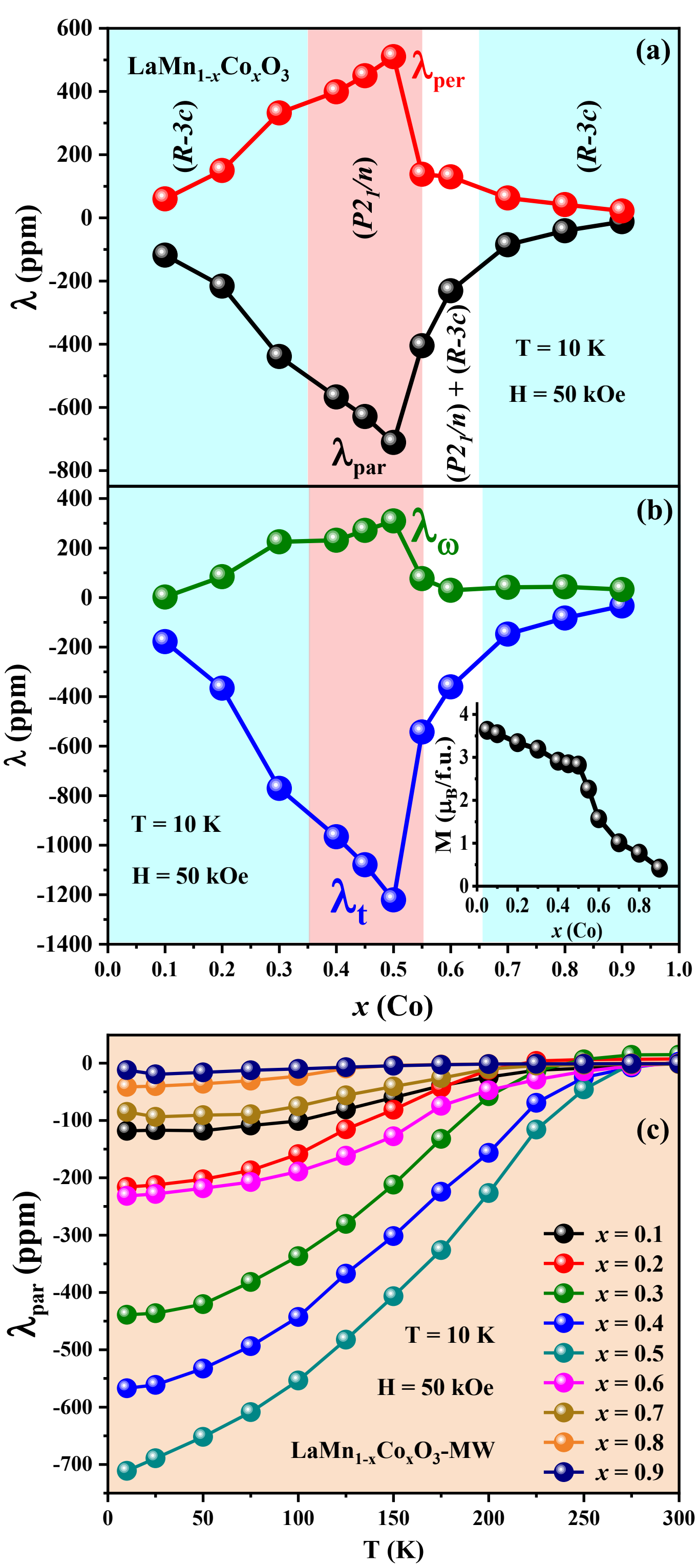

**Fig. 4.** Composition dependence of **(a)** longitudinal ($\lambda_{par}$), transverse ($\lambda_{per}$) magnetostrictions, **(b)** volume ($\lambda_{\omega}$) and anisotropic ($\lambda_t$) magnetostrictions and **(c)** Temperature dependence of magnetostriction measured at $T$ = 10 K under an applied magnetic field of $H$ = 50 kOe for $LaMn_{1-x}Co_xO_3$ ($x$ = 0.1–0.9) series. Inset in (b) reproduces the magnetization value at 50 kOe as a function of $x$.

We plot the composition dependence of magnetostrictions at 50 kOe extracted from isotherms at 10 K for $LaMn_{1-x}Co_xO_3$ in Fig. 4 (a) $\lambda_{par}$ & $\lambda_{per}$ and (b) $\lambda_t$ & $\lambda_{\omega}$. These graphs also include the magnetostriction values for compositions ($x$ = 0.1, 0.3, 0.45, 0.55, 0.7, and 0.9) that were not explicitly plotted in Fig. 3. The magnitudes of $\lambda_{par}$ and $\lambda_{per}$ vary strongly with Co content, and the variation is asymmetrical about $x$ = 0.5. $x$ = 0.5 exhibits maximum values of $\lambda_{par}$ = -711 ppm and $\lambda_{per}$ = 510 ppm. The decrease of $\lambda_{par}$ is steeper at $x > 0.5$ than for $x < 0.5$. The $\lambda_t$ and $\lambda_{\omega}$ also exhibit a similar trend to $\lambda_{par}$ and $\lambda_{per}$ as shown in Fig. 4(b). $\lambda_t$ ($\lambda_{\omega}$) = -1221 ppm (309 ppm) for $x$ = 0.5 decreases to -966 ppm (231 ppm) in $x$ = 0.4. On the right side of $x$ = 0.5, $\lambda_t$ ($\lambda_{\omega}$) = -543 ppm (76 ppm) at $x$ = 0.55 drastically decreases to -361 ppm (29 ppm) at $x$ = 0.6. Fig. 4(c) shows the temperature dependence of $\lambda_{par}$ values extracted at 50 kOe from field-dependent magnetostriction at different temperatures for $LaMn_{1-x}Co_xO_3$ for $x$ = 0.1-0.9. For all compositions, the magnitude of $\lambda_{par}$ is largest at 10 K and decreases gradually as temperature increases towards $T_C$ and becomes negligibly small above $T_C$. Magnetization and magnetostriction parameters for the $LaMn_{1-x}Co_xO_3$ series are presented in supplementary as Table. 2.

## D. Discussion

The two important observations are: 1. The anisotropic magnetostriction in $LaMn_{1-x}Co_xO_3$ increases with the Co content ($x$) from $x$ = 0.1 to 0.5 and then decreases abruptly around $x$ = 0.55-0.6 which is accompanied by a structural change from monoclinic for $x \leq 0.5$ to a mixed phase of monoclinic and rhombohedral ($x$ = 0.55-0.6) before reverting to purely rhombohedral for $x$ = 0.7. 2. The maximum anisotropic magnetostriction of $|\lambda_t|$ = 1221 ppm at 10 K and 50 kOe exhibited by $x$ = 0.5 is nearly twice the value found in $La_{0.5}Sr_{0.5}CoO_3$ ($\lambda_t$ = 620 ppm) at the same field, which contains $Co^{3+}$ and $Co^{4+}$ ions in equal proportion [23]. It is remarkable that $|\lambda_t|$ = 178 ppm (366 ppm) even for $x$ = 0.1 (0.2) which exceeds the typical value of 40-60 ppm found at 10 K in the ferromagnetic state of mixed valent manganites such as $La_{0.7}Ca_{0.3}MnO_3$ [39]. The presence of significant anisotropic magnetostriction even at low doping levels such as $x$ = 0.1 and 0.2 indicates that the Co ion is stabilized in a divalent state instead of a trivalent state. Based on XAS and resonance photoemission studies, J. -H. Park *et al*. [26] suggest that Co enters at Mn site in $LaMn_{0.85}Co_{0.15}O_3$ as $Co^{2+}(d^7)$ rather than anticipated $Co^{3+}$ ($d^6$). As a result, the fraction of $Mn^{3+}$ changes to $Mn^{4+}$ for charge compensation. $Co^{2+}(d^7{:}t_{2g}^5e_g^2)$ interacts ferromagnetically with neighboring $Mn^{4+}$ ($d^3{:}t_{2g}^3e_g^0$) via virtual hopping of $e_g$ electron. The ferromagnetically coupled $Co^{2+}$-$Mn^{4+}$ reaches the maximum density at $x = 0.5$. The ratio of orbital moment to spin magnetic moment in $LaMn_{0.5}Co_{0.5}O_3$ is $m_{orb}/m_{spin} = 0.47$ and orbital moment of $Mn^{4+}$ is negligible [4,5]. Hence, $Mn^{4+}$ is not responsible for the observed magnetostriction. The unquenched orbital moment associated with the degenerate $t_{2g}$ orbitals of $Co^{2+}$ couples to spin. As the spin magnetic moment rotates towards the direction of the applied field, the orbital moment also rotates and the atomic level distortion is transferred to the lattice via the crystal electric field, resulting in contraction or elongation of the sample length. The anisotropic magnetostriction reaches a maximum at $x = 0.5$, where the maximum density of $Co^{2+}$-$Mn^{4+}$ is realized. However, as $x$ increases above 0.5, the reverse charge transfer reaction $Co^{2+}$-$Mn^{4+} \rightarrow Co^{3+}$-$Mn^{3+}$ is activated in the rhombohedral phase. The sudden decrease of $\lambda_t$ for $x = 0.55$-$0.6$ implies a rapid decrease in $Co^{2+}$ content and an increase in $Co^{3+}$ in these compositions. Accordingly, $\lambda_t$ also decreases abruptly. The transformed $Co^{3+}$ ions may adopt a low-spin state ($S = 0$) and they are diamagnetic at low temperatures. Hence, ferromagnetic interaction between $Co^{2+}$-$Mn^{4+}$ pairs is disrupted. X-ray absorption spectroscopy shows that a small fraction of $Co^{2+}$ survives down to $x = 0.98$ while low-spin $Co^{3+}$ linearly increases with $x$ for $x > 0.6$ [27]. Further, antiferromagnetic interactions among like-charged ions ($Mn^{3+}$-$Mn^{3+}$ pairs and/or $Mn^{4+}$-$Mn^{4+}$ pairs) compete with weakened ferromagnetic interactions between remaining $Co^{2+}$-$Mn^{4+}$ pairs, resulting in magnetic frustration. As a result, long-range ferromagnetism is lost in the Co-rich compositions and $x > 0.55$ samples can be considered as cluster glass or spin glass. Hence, the magnitude of magnetostriction decreases as $x$ approaches 1.

## IV. Conclusion

In summary, $LaMn_{1-x}Co_xO_3$ ($x = 0.1$ to $0.9$) samples were synthesized by microwave irradiation of oxides. The structure changes from rhombohedral ($x \leq 0.3$) to monoclinic ($x = 0.4$-$0.55$) to rhombohedral ($x \geq 0.7$) and $x = 0.6$ shows a mixed phase. Anisotropic magnetostriction in $LaMn_{1-x}Co_xO_3$ at 10 K increases from 178 ppm for $x = 0.1$ to a maximum value of 1221 ppm for $x = 0.5$, then decreases abruptly for $x = 0.55$-$0.6$. Magnetic properties change abruptly around $x = 0.6$ suggesting a loss of long-range ferromagnetism. The evolution of anisotropic magnetostriction with composition indicates that ferromagnetically coupled $Co^{2+}$-$Mn^{4+}$ pairs increase in density as $x$ increases from 0.1 to 0.4, reach a maximum at $x = 0.5$, and decrease at high $x$ as the valence of Co cation shifts from +2 to +3. It will be interesting to theoretically understand why the anisotropic magnetostriction at 10 K in series is higher than Co-based spinels even though the spin-orbital coupling of $Co^{2+}$ cation makes the dominant contribution.

**ACKNOWLEDGEMENTS**

R. M. acknowledges the Ministry of Education, Singapore, for supporting this work (Grant no. A-8000924-00-000 and A-8001933-00-00).

**DATA AVAILABILITY**

The data that support the findings of this study are available from the corresponding author upon reasonable request.

**Supplemental:** Structural, magnetic, and magnetostriction parameters are tabulated.